\documentclass[runningheads]{llncs}
\usepackage{paralist}
\usepackage{units}
\usepackage{booktabs} % For formal tables
\usepackage{graphicx}
\usepackage{latexsym}
\usepackage{times}
\usepackage{multirow}
\usepackage{booktabs}
\usepackage{xspace}
\usepackage{cite}
\usepackage{subcaption}

\usepackage{todonotes}

\usepackage[all,pdftex]{xy}
\SelectTips{cm}{}

\newif\ifignore % when set to true, additional text appears containing
\ignorefalse

\renewcommand{\phi}{\varphi}

\newcommand{\R}{\mathbb{R}}

\newcommand{\Rpos}{\R_{+}}

\newcommand{\RN}{\R^{N}}
\newcommand{\RM}{\R^{M}}

\newcommand{\M}{\mathcal{M}}

\newcommand{\bv}{\mathbf{v}}
\newcommand{\bw}{\mathbf{w}}
\newcommand{\bu}{\mathbf{u}}
\newcommand{\sem}[1]{\llbracket #1 \rrbracket}

\newcommand{\DiaOp}[1]{\Diamond_{#1}}
\newcommand{\BoxOp}[1]{\Box_{#1}}
\newcommand{\speed}{\mathit{speed}}
\newcommand{\gear}{\mathit{gear}}
\newcommand{\rpm}{\mathit{rpm}}

\newcommand{\STL}{\textrm{STL}}

\newcommand{\Var}{\mathbf{Var}}

\newcommand{\UntilOp}[1]{\mathbin{\mathcal{U}_{#1}}}

\newcommand{\Rnn}{\R_{\ge 0}}

\newcommand{\Defeq}{:=}
\newcommand{\Robust}[2]{{ \llbracket #1, #2 \rrbracket}}

\newcommand{\Vee}[1]{{{\bigsqcup_{#1}}}}
\newcommand{\Wedge}[1]{{{\bigsqcap_{#1}}}}
\newcommand{\throttle}{\mathit{throttle}}
\newcommand{\brake}{\mathit{brake}}

\newcommand{\breach}{\textsc{Breach}\xspace}
\newcommand{\sTaliro}{\textsc{S-Taliro}\xspace}
\newcommand{\FalStar}{\textsc{FalStar}\xspace}

\newcommand{\fit}{\mathit{GCF}}
\newcommand{\fitfal}{\mathit{GCF}_{\mathsf{fal}}}
\newcommand{\fitfalsf}{\mathit{GCF}_{\mathsf{fal\_sf}}}
\newcommand{\Rmax}[1]{\ensuremath{\mathit{R_{max}^{#1}}}\xspace}

\usepackage{mathtools}
\DeclarePairedDelimiter{\ceil}{\lceil}{\rceil}

\newcommand{\bx}{\mathbf{x}}

\newcommand{\trans}{\ensuremath{\mathit{T}}\xspace}
\spnewtheorem{mytheorem}{Theorem}%[section]
{\bfseries}{\itshape} 
\spnewtheorem{mylemma}[mytheorem]{Lemma}{\bfseries}{\itshape}
\spnewtheorem{myproposition}[mytheorem]{Proposition}{\bfseries}{\itshape}
\spnewtheorem{mysublemma}[mytheorem]{Sublemma}{\bfseries}{\itshape}
\spnewtheorem{mycorollary}[mytheorem]{Corollary}{\bfseries}{\itshape}
\spnewtheorem{myfact}[mytheorem]{Fact}{\bfseries}{\itshape}
\spnewtheorem{mynotation}[mytheorem]{Notation}{\bfseries}{\rmfamily}
\spnewtheorem{myremark}[mytheorem]{Remark}{\bfseries}{\rmfamily}
\spnewtheorem{myexample}[mytheorem]{Example}{\bfseries}{\rmfamily}
\spnewtheorem{myassumption}[mytheorem]{Assumption}{\bfseries}{\rmfamily}
\spnewtheorem{mydefinition}[mytheorem]{Definition}{\bfseries}{\rmfamily}
\spnewtheorem{myrequirements}[mytheorem]{Requirements}{\bfseries}{\rmfamily}
\spnewtheorem{myproblem}[mytheorem]{Problem}{\bfseries}{\rmfamily}
\spnewtheorem{myinstance}[mytheorem]{Instance}{\bfseries}{\rmfamily}

\usepackage{bm}
\usepackage{amsmath}
\usepackage{amssymb}
\usepackage{amsfonts}

\usepackage{stmaryrd}
\usepackage{graphicx}
\usepackage{color}
\usepackage{colortbl}

\usepackage{proof}

\usepackage{wrapfig}

\usepackage{tabularx}
\usepackage[misc]{ifsym}

\usepackage{arydshln}
\usepackage{hyperref}
\hypersetup{
colorlinks = false, % false: boxed links; true: colored links
hidelinks = true,
linkcolor=black, % color of internal links
citecolor=black, % color of links to bibliography
urlcolor=black, % color of external links
filecolor=black
}

\usepackage{enumitem}

\usepackage{multirow}
\usepackage{pifont}

\newcounter{researchquestionCount}
\newcommand{\researchquestion}[1]{\stepcounter{researchquestionCount}\vspace{10pt}\noindent\parbox{0.97\textwidth}{{\bf RQ\arabic{researchquestionCount}} {\it #1}}\vspace{0pt}}

\newcommand{\pos}{\mathit{Pos}}
\newcommand{\reff}{\mathit{Ref}}
\newcommand{\closeref}{\mathit{close\_ref}}
\newcommand{\reachref}{\mathit{reach\_ref\_in\_tau}}
\newcommand{\nnreq}{\mathit{NN\_req}}
\newcommand{\pedalangle}{\mathit{Pedal\_Angle}}
\newcommand{\enginespeed}{\mathit{Engine\_Speed}}

\newcommand{\BA}{\ensuremath{\mathtt{BA}}\xspace}
\newcommand{\constrEmb}{\ensuremath{\mathtt{CE}}\xspace}
\newcommand{\lmBasic}{\ensuremath{\mathtt{LM}}\xspace}
\newcommand{\lmAdv}{\ensuremath{\mathtt{LM_{sf}}}\xspace}

\begin{document}
\title{Constraining Counterexamples in Hybrid System Falsification: Penalty-Based Approaches\thanks{The authors are supported by ERATO HASUO Metamathematics for Systems Design Project (No. JPMJER1603), JST; Zhenya Zhang is supported by Grant-in-Aid for JSPS Fellows No. 19J15218.}}
\titlerunning{Constraining Counterexamples in Hybrid System Falsification}

\author{Zhenya Zhang\inst{1,2,3}\textsuperscript{(\Letter)}\orcidID{0000-0002-3854-9846}
\and
Paolo Arcaini\inst{1}\orcidID{0000-0002-6253-4062}
\and
Ichiro Hasuo\inst{1,2}\orcidID{0000-0002-8300-4650}}

\authorrunning{Z. Zhang et al.}

\institute{National Institute of Informatics, Tokyo, Japan\\
\email{\{zhangzy,arcaini,hasuo\}@nii.ac.jp}
\and
SOKENDAI (The Graduate University for Advanced Studies), Hayama, Japan
\and JSPS Research Fellow, Tokyo,  Japan
}

\maketitle

\begin{abstract}
\emph{Falsification} of hybrid systems is attracting ever-growing attention in quality assurance of Cyber-Physical Systems (CPS) as a practical alternative to exhaustive formal verification. In falsification, one searches for a falsifying input that drives a given black-box model to output an undesired signal. In this paper, we identify \emph{input constraints}---such as the constraint ``the throttle and brake pedals should not be pressed simultaneously'' for an automotive powertrain model---as a key factor for the practical value of falsification methods. We propose three approaches for systematically addressing input constraints in optimization-based falsification, two among which come from the lexicographic method studied in the context of constrained multi-objective optimization. Our experiments show the approaches' effectiveness. 
\keywords{Hybrid System Falsification, Signal Temporal Logic, Constraints, Penalty, Lexicographic methods}
\end{abstract}

\section{Introduction}

Cyber-physical systems (CPS) combine physical systems with digital controllers: while the former are characterized by continuous dynamics, the latter are inherently discrete. Such a combination is usually named as \emph{hybrid systems}. The continuous dynamics of hybrid systems leads to infinite search spaces, and this makes their formal verification---especially automated methods---almost impossible. Therefore, research has followed a more pragmatic approach by pursuing the \emph{falsification} of the system: since checking whether all inputs satisfy the specification is not feasible, falsification considers the opposite problem and looks for an input that violates it. Formally, given a \emph{model} $\mathcal{M}$ that takes an input signal $\bu$ and outputs a signal $\mathcal{M}(\bu)$, and a \emph{specification} $\varphi$ (a temporal formula), the falsification problem consists in finding a \emph{falsifying input}, i.e., an input signal $\bu$ such that the corresponding output $\mathcal{M}(\bu)$ violates $\varphi$. 

State-of-the-art falsification approaches see the falsification problem as an optimization problem. This is possible thanks to the \emph{robust semantics} of temporal formulas~\cite{FainekosP09,DonzeM10}; instead of the classical Boolean satisfaction relation $\bv\models\varphi$, robust semantics assigns a value $\sem{\bv, \varphi} \in\R \cup \{\infty,-\infty\}$ that assesses not only whether $\varphi$ is satisfied or violated (by the sign), but also \emph{how robustly} the formula is satisfied or violated. Falsification algorithms exploit this fact by iteratively generating inputs in the direction of decreasing robustness, with the aim of finding an input with negative robustness (i.e., a falsifying input). Different optimization-based falsification algorithms have been developed~\cite{FainekosP09,Annpureddy-et-al2011,AdimoolamDDKJ17,DeshmukhJKM15,KuratkoR14,Donze10,DonzeM10,DreossiDDKJD15,ZutshiDSK14,0001SDKJ15,SilvettiPB17,falsificationTCAD2018,AkazakiLYDH18,ernstQEST2019}. See~\cite{KapinskiDJIB16} for a survey. Moreover, also tools have been developed, as \breach~\cite{Donze10}, \sTaliro~\cite{Annpureddy-et-al2011}, and \FalStar~\cite{falsificationTCAD2018}, that work with Simulink models.

In real scenarios, there usually exist some \emph{(input) constraints} $\psi$ over input signals. For example, in an automotive system, one usually assumes that \emph{throttle} and \emph{brake} should not be positive at the same time. Descriptions of CPS sometimes report constraints on the system inputs, e.g., \cite{JinDKUB14,Abdessalem2018TVC}. Therefore, when generating inputs for the falsification problem, we should also guarantee that those inputs respect the constraints; otherwise, there is the risk that the resulting falsifying input is unrealistic and thus useless. However, not too many research efforts have been spent on this problem in the falsification community. To the best of our knowledge, explicit attempts to consider input constraints in falsification have been made only in~\cite{BarbotNFM2019}. In~\cite{BarbotNFM2019},  constraints are represented in terms of a timed automaton, and inputs to be used for falsification are sampled from the accepted words of the automaton. The main drawback of this approach is that it can only rely on sampling for falsification, and cannot take advantage of more efficient optimization-based techniques. 

\paragraph{Contribution}
In this paper we propose three approaches in which input constraints are addressed explicitly in the falsification problem, and that still benefit from optimization-based techniques. The general idea of the three approaches is to add a penalty factor to the objective function for the inputs that do not satisfy the input constraints.

The first proposed approach consists in modifying the specification under falsification in $\psi \to \varphi$: the only way to falsify the whole formula is to satisfy the input constraint $\psi$ and falsify the specification $\varphi$. The penalty factor for the violation of the input constraints is directly given by the STL robustness.

Our second approach employs the \emph{lexicographic method}, a method developed in multi-objective optimization~\cite{Chang2015}. In our adaptation of the method, the satisfaction of the input constraints is embedded in a \emph{global cost function} that must be minimized: if the input constraints are not satisfied,  the cost function is principally determined by the \emph{degree of violation} of the input constraints. In contrast, if the input constraints are satisfied, the cost function value is only determined by the robustness of the specification (as in the classical unconstrained falsification setting). The advantage of the approach is that the satisfaction of the input constraints is prioritized w.r.t.\ the falsification of the specification: indeed, it is useless to find a falsifying input that does not respect the constraints.

The third approach tries to improve the second approach by simulating the model only when the input constraints are satisfied. Although this can reduce the accuracy of the search, it can also speed up the falsification process.

The three approaches have been experimented over 3 Simulink models and 17 specifications that are used in falsification competitions~\cite{ARCHCOMP19Falsification}; for each model, we experimented the approaches using several input constraints of different complexity. Experimental results show that the approaches can effectively handle the constraints. In terms of falsification capability, no approach is strictly better than the others, although lexicographic methods seem better on average.

\paragraph{Paper structure}
\S{}\ref{sec:background} introduces some necessary background on the kind of models, specifications, and algorithms used in falsification. Then, \S{}\ref{sec:proposedApproach} presents our proposed approach, and \S{}\ref{sec:experiments} describes some experiments we performed to evaluate it. Finally, \S{}\ref{sec:related} reviews some related work, and \S{}\ref{sec:conclusion} concludes the paper.

\section{Background}\label{sec:background}

In this section, we review the widely-accepted method of hill-climbing optimization-based falsification. The core of making use of hill-climbing optimization is the introduction of \emph{robust semantics} of temporal formulas.

\subsection{Robust Semantics for STL}
Our definitions here are taken from~\cite{FainekosP09,DonzeM10}. 

\begin{mydefinition}[(Time-bounded) signal]\label{def:timeBoundedSignal}
Let $T\in \Rpos$ be a positive real. An \emph{$M$-di\-men\-sion\-al signal}  with a time horizon $T$ is a function $\bw\colon [0,T]\to\R^{M}$.

Let $\bw\colon [0,T]\to \RM$ and $\bw'\colon [0,T']\to\RM$ be $M$-dimensional signals. Their \emph{concatenation} $\bw\cdot\bw'\colon [0,T+T']\to \RM$ is the $M$-dimensional signal defined by
\begin{math}
(\bw\cdot\bw')(t) = \bw(t)
\end{math}
if $t\in [0,T]$, and $ (\bw\cdot\bw')(t)=\bw'(t-T)$ if $t\in(T,T+T']$.

Let $0<T_{1}<T_{2}\le T$. The \emph{restriction} $\bw|_{[T_{1},T_{2}]}\colon [0,T_{2}-T_{1}]\to \RM$ of $\bw\colon [0,T]\to \RM$ to the interval $[T_{1},T_{2}]$ is defined by $(\bw|_{[T_{1},T_{2}]})(t)=\bw(T_{1}+t)$. 
\end{mydefinition}
We treat the system model as a black box, i.e., the system behaviors are only observed from inputs and their corresponding outputs. We therefore simply define the system model as a function.
\begin{mydefinition}[System model $\M$]\label{def:systemModel}
A \emph{system model}, with $M$-dimensional input and $N$-dimensional output, is a function $\mathcal{M}$ that takes an input signal $\bu\colon [0,T]\to \R^{M}$ and returns a signal $\mathcal{M}(\bu)\colon [0,T]\to \R^{N}$. Here the common time horizon $T\in \Rpos$ is arbitrary. Furthermore, we impose the following \emph{causality} condition on $\mathcal{M}$: for any time-bounded signals $\bu\colon [0,T]\to \R^{M}$ and $\bu'\colon [0,T']\to \R^{M}$, we require that
\begin{math}\label{eq:causality}
\mathcal{M}(\bu\cdot\bu') \big|_{[0,T]} = \mathcal{M}(\bu)
\end{math}.
\end{mydefinition}

\begin{mydefinition}[STL syntax]\label{def:stlSyntax}
We fix a set $\Var$ of variables. In $\STL$, \emph{atomic propositions} and \emph{formulas} are defined as follows, respectively:
\begin{math}
%  \AP \ni 
\alpha 
\,::\equiv\,
f(x_1, \dots, x_N) > 0
\end{math}, and 
\begin{math}
%  \Fml \ni 
\varphi \,::\equiv\,
\alpha \mid \bot
% \mid \top
\mid \neg \varphi 
\mid \varphi \wedge \varphi 
\mid \varphi \vee \varphi 
% \mid \varphi \wedge \varphi
\mid \varphi \UntilOp{I} \varphi
% \mid \varphi \TUntil{[a,b]} \varphi 
% \mid \varphi \LTUntil{[a,b]} \varphi 
\end{math}. Here
$f$ is an $N$-ary function $f:\RN \to \R$, $x_1, \dots, x_N \in \Var$,
% $a,b \in \Rnn$ such that $a<b$,
and $I$ is a closed non-singular interval in $\Rnn$,
i.e.\ $I=[a,b]$ or $[a, \infty)$ where $a,b \in \R$ and $a<b$.
%\todo{In Def. 3 (page 3), why do you have to exclude point intervals?}
% Eventually operator $\DiaOp{I}\varphi$ can be derived by $\DiaOp{I}{\varphi} ::\equiv \top\UntilOp{I} \varphi$; and always operator $\BoxOp{I}{\varphi}$
% can be derived by $\BoxOp{I}{\varphi} ::\equiv \neg \DiaOp{I}{\neg\varphi}$.
\end{mydefinition}
We omit subscripts $I$ for temporal operators if $I = [0, \infty)$. Other common connectives such as $\rightarrow,\top$, $\Box_{I}$ (always) and $\Diamond_{I}$ (eventually), are introduced as abbreviations: $\varphi_1 \rightarrow \varphi_2 \equiv \neg \varphi_1 \vee \varphi_2$, $\Diamond_{I}\varphi\equiv\top\UntilOp{I}\varphi$ and 
$\Box_{I}\varphi\equiv\lnot\Diamond_{I}\lnot\varphi$. An atomic formula $f(\vec{x})\le c$, where $c\in\R$, is accommodated using $\lnot$ and the function $f'(\vec{x}):=f(\vec{x})-c$.

%The definition of STL robustness semantics 
\begin{mydefinition}[Robust semantics~\cite{DonzeM10}]\label{def:robSemantics} 
Let $\bw \colon [0,T]\to \R^{N}$ be an $N$-dimensional signal, and $t\in [0,T)$. The \emph{$t$-shift} of $\bw$, denoted by $\bw^t$, is the time-bounded signal $\bw^t\colon [0,T-t]\to \R^{N}$ defined by $\bw^t(t') \Defeq \bw(t+t')$.

Let $\bw \colon [0,T] \to \R^{|\Var|}$ be a signal, and $\varphi$ be an $\STL$ formula. We define the \emph{robustness} $\Robust{\bw}{\varphi} \in \R \cup \{\infty,-\infty\}$ as follows, by induction on the construction of formulas. Here $\bigsqcap$ and $\bigsqcup$ denote infimums and supremums of real numbers, respectively. Their binary version $\sqcap$ and $\sqcup$ denote minimum and maximum.
\begin{align*}
&\Robust{\bw}{f(x_1, \cdots, x_n) > 0} \;\Defeq \;
f\bigl(\bw(0)(x_1), \cdots, \bw(0)(x_n)\bigr) 
\\
&\Robust{\bw}{\bot} \;\Defeq\; -\infty
\qquad
\Robust{\bw}{\neg \varphi}  \;\Defeq\;  - \Robust{\bw}{\varphi}\qquad 
\\
&
\Robust{\bw}{\varphi_1 \wedge \varphi_2}  \;\Defeq\;  \Robust{\bw}{\varphi_1} \sqcap \Robust{\bw}{\varphi_2}
\qquad
\Robust{\bw}{\varphi_1 \lor \varphi_2}   \;\Defeq\;   \Robust{\bw}{\varphi_1} \sqcup \Robust{\bw}{\varphi_2}
\\
&
\Robust{\bw}{\varphi_1 \UntilOp{I} \varphi_2}   \;\Defeq\; 
             \textstyle{ \Vee{t \in I\cap [0,T]}\bigl(\,\Robust{\bw^t}{\varphi_2} \sqcap 
             \Wedge{t' \in [0, t)} \Robust{\bw^{t'}}{\varphi_1}\,\bigr)}
\end{align*}
\end{mydefinition}

For atomic formulas, $\Robust{\bw}{f(\vec{x})>c}$ stands for the vertical margin $f(\vec{x})-c$ for the signal $\bw$ at time $0$. A negative robustness value indicates how far the formula is from being true. It follows from the definition that the robustness for the eventually modality is given by
\begin{math}
\Robust{\bw}{\DiaOp{[a,b]} (x > 0)}
= \Vee{t \in [a,b]\cap[0,T]} \bw(t)(x)
\end{math}. 

The above robustness notion taken from~\cite{DonzeM10} is therefore \emph{spatial}. Other robustness notions take \emph{temporal} aspects into account, too, such as ``how long before the deadline the required event occurs.'' See e.g.~\cite{DonzeM10,AkazakiH15}. Our choice of spatial robustness in this paper is for the sake of simplicity, and is thus not essential.

The original semantics of $\STL$ is Boolean, given as usual by a binary relation $\models$ between signals and formulas. The robust semantics refines the Boolean one in the following sense:
$ \sem{\bw,\varphi} > 0$
implies
$\bw\models\varphi$, and
$ \sem{\bw,\varphi} < 0$
implies
$\bw\not\models\varphi$, see~\cite[Prop.~16]{FainekosP09}.
Optimization-based falsification via robust semantics hinges on this refinement. 
% Although the definitions so far are for time-unbounded signals,
% we note that the robust semantics $\sem{\bw,\varphi}$,
% as well as the Boolean satisfaction $\bw\models\varphi$, can be easily adapted to time-bounded signals (Def.~\ref{def:timeBoundedSignal}).

\subsection{Hill Climbing-Guided Falsification}\label{subsec:hillClimbingFalsification}
For the falsification problem, hill-climbing optimization is the main applied technique~\cite{Annpureddy-et-al2011,AdimoolamDDKJ17,DeshmukhJKM15,KuratkoR14,Donze10,DonzeM10,DreossiDDKJD15,ZutshiDSK14,AkazakiKH17,SilvettiPB17,DreossiDS17,KapinskiDJIB16}, and different tools exist, as \breach~\cite{Donze10}, \sTaliro~\cite{Annpureddy-et-al2011}, and \FalStar~\cite{falsificationTCAD2018}. We here formulate the falsification problem.
%that will be used later in the proposed approach.

\begin{mydefinition}[Falsifying input]\label{def:falsifyingInput}
Let $\mathcal{M}$ be a system model, and~$\varphi$ be an STL formula. A signal $\bu\colon [0,T]\to \R^{|\Var|}$ is a \emph{falsifying input} if $\Robust{\mathcal{M}(\bu)}{\varphi} < 0$; the latter implies $\mathcal{M}(\bu)\not\models\varphi$.
\end{mydefinition}

\begin{wrapfigure}[11]{r}{0.4\textwidth}
\includegraphics[width=0.38\textwidth]{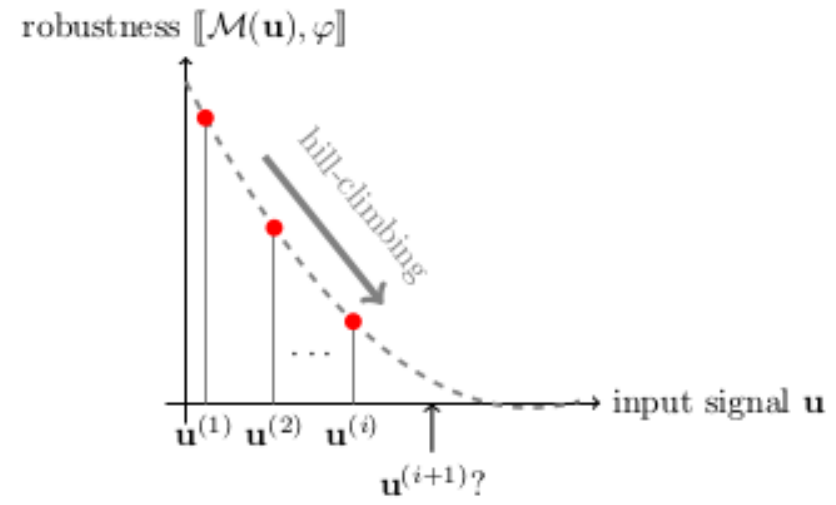}
\caption{Hill-climbing optimization}
\label{fig:hillClimbing}
\end{wrapfigure}

%bounded optimization problem derived from falsification
\begin{mydefinition}[Unconstrained falsification problem]\label{def:unconstrFals}
The technique for solving a falsification problem is via transforming it into an optimization problem, shown as follows: 
\begin{equation}
\begin{aligned}
& \underset{\bu}{\text{minimize}}
& & \Robust{\M(\bu)}{\varphi} \\
& \text{subject to}
& & \bu \in \Omega
\end{aligned}
\end{equation}
\end{mydefinition}
In practice, the system input signal $\bu$ is represented with a finite set of variables defined over the search space $\Omega$ (a hyperrectangle)\footnote{Although the  problem  has  a simple form of constraints, we prefer to name it \emph{unconstrained} to  distinguish it from the constrained setting we introduce later.}. The use of quantitative robust semantics $\Robust{\mathcal{M}(\bu)}{\varphi}\in \R\cup \{\infty,-\infty\}$ in the above problem enables the use of hill-climbing optimization. Hill climbing is a family of metaheuristics-based optimization algorithms, which is usually used for handling black-box optimization. The hill-climbing optimization scheme is shown in Fig.~\ref{fig:hillClimbing}.

The algorithm is iterative: in every loop, it takes some samplings and computes the fitness of them. Globally, the sampling process is divided into two stages: random initial samplings and the sequent samplings based on the observation of sampling history in order to minimize the objective function. The most expensive step in each loop is given by the computation of the fitness that requires to simulate the system. Hill-climbing includes various implementations of stochastic optimization algorithms. Examples are CMA-ES~\cite{AugerH05} (used in our experiments), SA, and GNM~\cite{LuersonLeRiche2004}.

%\section{STL propagation (inference): a baseline approach}
%\subsection{Interactive methods: making use of STL robust semantics}
%In lexicographic methods, all the inputs that violate the constraints have a higher fitness than those satisfying ones, so they are not considered during the search process; in comparison, interactive methods take a milder strategy: it keeps interacting with users during the search by continuously requesting preference information and making decisions accordingly. As a result, although those violating inputs will not constitute the final solution, they contribute insights into the direction where search should proceed.

%In our problem setting, interactive methods can be achieved with a simple approach. We transform the problem of Def.~\ref{} into a single-objective optimization problem in the following way:

\section{Penalty-Based Approaches for Handling Input Constraints}\label{sec:proposedApproach}

The problem setting of falsification introduced in Def.~\ref{def:unconstrFals} does not take into consideration possible constraints over the input signals, which limits the practicality of the falsification techniques in real contexts. Indeed, some works~\cite{JinDKUB14,Abdessalem2018TVC} report that input constraints do exist in CPS.

In this paper, we tackle the problem of handling input constraints in optimization-based falsification.

\begin{mydefinition}[Constrained falsification problem]\label{def:constrOptProblem}
The constrained falsification problem can be stated as follows, where $\psi$ are input constraints, expressed in STL, over the input signals $\bu$.
\begin{equation*}
\begin{aligned}
& \underset{\bu}{\text{minimize}}
& & \Robust{\M(\bu)}{\varphi} \\
& \text{subject to}
& & \bu \models \psi\\
%& & \Robust{\bu}{\psi} > 0\\
& & & \bu \in \Omega
\end{aligned}
\end{equation*}
\end{mydefinition}

In our approach, input constraints $\psi$ are assumed to be expressible in STL.
% , which is expressive enough to specify any reasonable constraint over the input signals $\bu$.

More generally, constraints in optimization solutions have been studied in the field of optimization: see~\cite{Zhun2017constrSearch} for an overview.
For example, the \emph{death penalty} method~\cite{BackHS91} discards all the solutions that violate the constraints; while this method can work well when the feasible search space is convex, it does not work well in general, and particularly in our context where the constraints can be arbitrarily complex. Other more advanced methods (static penalty, dynamic penalty, and adaptive penalty) add a penalty factor to the objective function~\cite{Zhun2017constrSearch}, so that solutions violating the constraints are penalized during the search. 

In this work, we follow this second line of research in which we add, to the objective function of falsification, a penalty related to the non-satisfaction of the input constraints. We propose three approaches: a simple approach based on the modification of the specification under study is presented in \S{}\ref{sec:stlBasedApproach}, while two more advanced approaches based on lexicographic methods are proposed in \S{}\ref{sec:lexMethodFals} and \S{}\ref{sec:lexMethodFalsVar}.

\subsection{Constraint Embedding approach}\label{sec:stlBasedApproach}

A straightforward penalty-based approach to the constrained falsification problem consists in embedding the input constraints $\psi$ as a prerequisite of the system specification $\varphi$. In this way, we obtain the STL formula $\psi\to\varphi$ as a new falsification goal.

The constrained problem of Def.~\ref{def:constrOptProblem} can be stated as the following unconstrained problem.

\begin{equation*}%\label{eq:constrInFormula}
\begin{aligned}
& \underset{\bu}{\text{minimize}}
& & \Robust{\langle\bu, \M(\bu)\rangle}{\psi\to\varphi} \\
& & & \bu \in \Omega
\end{aligned}
\end{equation*}

The falsification approach must now evaluate the robustness of a formula that predicates both over the input and output signals, formally denoted as $\langle\bu, \M(\bu)\rangle$.

The soundness of the approach is given by Thm.~\ref{thm:soundConstrEmb}.

\begin{theorem}[Soundness and completeness of the Constraint Embedding Approach]\label{thm:soundConstrEmb}
	%\todo{Theorem 1 can be strengthened: iff}
%If there exists an input signal $\bu$ such that $\Robust{\langle\bu,\M(\bu)\rangle}{\psi\to\varphi} \le 0$, then the input constraints $\psi$ are satisfied and the specification $\varphi$ is falsified.
For all input signals $\bu$,
$\Robust{\langle\bu,\M(\bu)\rangle}{\psi\to\varphi} < 0$
if and only if
the input constraints $\psi$ are satisfied and the specification $\varphi$ is falsified.
\end{theorem}

The proof directly comes from the robustness definition of STL and the semantics of the implication.

%\subsection{The lexicographic method}\label{sec:lexMethodBack}
\subsection{Lexicographic Method approach}\label{sec:lexMethodFals}

%While these techniques can be very effective, in general they do not provide any guarantee regarding the satisfaction of the constraints. In order to provide some guarantee, the penalties should impose an \emph{order} between the satisfaction of the constraints and the optimization of the objective function(s). This is exactly the aim of the \emph{lexicographic methods}~\cite{Chang2015} that we employ in this work.

While the constraint embedding approach can be effective in some cases, it does not dictate a search algorithm to first satisfy input constraints $\psi$ and then falsify the specification $\varphi$. We here propose a method that imposes a strict prioritization between the satisfaction of the input constraints and the optimization of the objective function for falsification. This method is based on the use of a \emph{lexicographic method}~\cite{Chang2015} for defining the fitness function of the optimization problem.

%Then, in \S{}\ref{sec:lexMethodFals} and \S{}\ref{sec:lexMethodFalsVar}, we propose two versions of the technique we propose (based on a lexicographic method) to solve the problem stated in Def.~\ref{def:constrOptProblem}.

A lexicographic method~\cite{Chang2015} can be applied for a multi-objective optimization problem that aims at minimizing objective functions $f_1, \ldots, f_N$, and for which there exists a preference order in the optimization of the objective functions, i.e., functions with higher priorities must be optimized first. Formally, there exists a total order of priorities $p_1,$ $\ldots,$ $p_N$, where $p_k = N-k$ for each $k\in\{1,\ldots, N\}$; the larger $p_k$ is, the higher priority $f_k$ has.
\begin{align}\label{eq:objFunctionsLexi}
& \underset{\bx}{\text{minimize}}
& & f_1(\bx), \dots, f_N(\bx)\\ 
& \text{subject to}
& & \bx \in \Omega \nonumber
\end{align}
The method defines a global cost function $\fit$ in the following way:
\begin{equation}\label{eq:generalGFC}
\fit(\bx) = \sum_{k = 1}^{N}{B^{p_k} \ceil{(B-1) \trans_k\bigl(f_k(\bx)\bigr)}}
\end{equation}
where $B\in\Rpos$ with $B>1$ is a base number, $\ceil{}$ is the regular ceiling operator, and each $\trans_k$ is a transformation function. Note that $\ceil{(B-1) \trans_k\bigl( f_k(\bx)\bigr)}$ is needed to map the transformed value of the objective function $f_k$ in $B$ quantization levels. Such a quantization is required by the lexicographic method to maintain the total order of the inputs~\cite{Ehrgott2005} w.r.t. the priorities of the objective functions, i.e., the fitness value of a unachieved function with higher priority always dominates the fitness values of functions of lower priority. Note that the value of $B$ can have an effect on the efficiency of the search~\cite{Ehrgott2005}, as also noted during the application of the lexicographic methods in other contexts~\cite{pinchera2017lexicographic}. In the experiments, we will evaluate such effect using different values for $B$.

The definition of a $\trans_k$ is specific to the type of optimization problem; for example, we will see later how to define it for the constraint satisfaction problem and the falsification problem. In any case, the definition of a $\trans_k$ must at least satisfy the monotonicity property, i.e., given two values $v_1 \le v_2$, then $\trans_k(v_1)\le \trans_k(v_2)$. Usually, a transformation function $\trans_k$ is implemented as a normalization function between [0,1]: in such a case, the values of $f_k$ that are mapped to 0 are those that \emph{achieve} the objective.\footnote{Note that, in general, it is not always possible to specify when an objective function is ``achieved''. However, the lexicographic methods require that for functions $f_1, \ldots, f_{N-1}$, this is possible, and this is applicable in our context.}

We apply the lexicographic method to the constrained falsification problem introduced in Def.~\ref{def:constrOptProblem}. To do this, we first turn the constrained falsification problem in a unconstrained multi-objective problem as follows.
\begin{align}
& \underset{\bu}{\text{minimize}}
& & \Robust{\bu}{\neg\psi}\label{eq:constraint}\\
& \underset{\bu}{\text{minimize}}
& & \Robust{\M(\bu)}{\varphi}\label{eq:robustness}\\
& \text{subject to}
& & \bu \in \Omega \nonumber
\end{align}
The constraint satisfaction problem has been turned into an optimization problem by exploiting the robust semantics of STL (recall that also the input constraints are expressed in STL). Since in a lexicographic method all objective functions must be minimized (see Eq.~\ref{eq:objFunctionsLexi}), we consider the negation of the input constraints (negative robustness of $\neg \psi$ corresponds to positive robustness of $\psi$).

We can now combine the two objectives (Eq.~\ref{eq:constraint} and Eq.~\ref{eq:robustness}) into a single global cost function, following Eq.~\ref{eq:generalGFC}. Since we want to prioritize the satisfaction of the input constraints, we take $\Robust{\bu}{\neg\psi}$ as $f_1$, and $\Robust{\M(\bu)}{\varphi}$ as $f_2$. The definition of the global cost function is as follows.

\begin{mydefinition}[Lexicographic fitness function $\fitfal$ for falsification]\label{def:lexicographicForFals}
Let $f_1(\bu) \Defeq \Robust{\bu}{\neg\psi}$, and $f_2(\bu)\Defeq \Robust{\M(\bu)}{\varphi}$. The definition of the global cost function for the constrained falsification problem is as follows:
\[\fitfal(\bu) = B\ceil{(B-1) \trans_1(f_1(\bu))} + {(B-1) \trans_2(f_2(\bu))}\]
\end{mydefinition}

As explained before, the definition of a transformation function $\trans_k$ is specific to the kind of optimization problem. In our context, the transformation function $\trans_1$ considers values $r$ given by the robustness evaluation of the input constraints: for any negative value of the robustness, the input constraints are satisfied, while positive values indicate the degree of violation of the input constraints $\psi$. Therefore, $\trans_1$ is defined as a normalization function as follows:
\begin{equation}\label{eq:defTconstr}
\trans_1(r)=\left\{
\begin{array}{lll}
0 & & r < 0\\
\dfrac{r}{\Rmax{\psi}}& & \text{otherwise}\\
\end{array}
\right.
\end{equation}
where \Rmax{\psi} is the possible maximum value of $r$. The identification of a correct \Rmax{\psi} requires minimum effort by sampling the input space. We will present how we come up with \Rmax{\psi} later in \S{}\ref{sec:experiments}.

%The transformation function $\trans_2$, instead, considers values $r$ given by the robustness evaluation of the specification $\varphi$. Also in this case, negative values of the robustness mean that the objective is achieved (i.e., the specification is falsified); however, differently from the constraints satisfaction problem, here we are also interested in maintaining the robustness value when the objective is achieved (i.e., \emph{how much} the specification is falsified). Therefore, the definition of the transformation function for $\trans_2$ is as follows:
The transformation function $\trans_2$, instead, considers values $r$ given by the robustness evaluation of the specification $\varphi$. Also in this case, negative values of the robustness mean that the objective is achieved (i.e., the specification is falsified). Therefore, the definition of the transformation function for $\trans_2$ is as follows:
\begin{equation}\label{eq:defTspec}
\trans_2(r)=\left\{
\begin{array}{lll}
0 & & r < 0\\
\epsilon&&r = 0\\
\dfrac{r}{\Rmax{\varphi}}& & \text{otherwise}\\
\end{array}
\right.
\end{equation}
where \Rmax{\varphi} is the possible maximum value of $r$, and $\epsilon$ is an arbitarily small positive number\footnote{Note that this is needed to  distinguish   inputs having robustness 0 (not falsifying) from those having negative robustness (falsifying).}. We will also explain later in \S{}\ref{sec:experiments} how we select a proper \Rmax{\varphi}.

Considering the definitions of the two transformation functions, we can now analyse the behaviour of function $\fitfal$ (see Def.~\ref{def:lexicographicForFals}). Given an input signal $\bu$, if the input constraints $\psi$ are satisfied, the first operand of the sum will be 0 (due to the transformation function $\trans_1$ in Eq.~\ref{eq:defTconstr}), and therefore the value of $\fitfal$ will only depend on the robustness value of the temporal specification (i.e., the second operand). On the other hand, if the input constraints are not satisfied, the first operand will be positive and guaranteed to be larger than the second one (so driving the search towards the satisfaction of the input constraints).

Note that in the definition of $\fitfal$, we do not apply the ceiling operator to the robustness evaluation of the specification $\varphi$ (i.e., $f_2$). It is indeed known that the ceiling operator is not really needed by the lexicographic method for the last operand of the sum~\cite{Chang2015,pinchera2017lexicographic}, and we take advantage of this. Therefore, since $f_2$ corresponds to the falsification algorithm, we prefer to remove the ceiling in order to preserve as much information as possible regarding the specification robustness that could be helpful for driving the search. Indeed, removing the ceiling avoids the quantization effect that in general is adversarial for the hill-climbing search.

\begin{theorem}[Soundness of the $\fitfal$ fitness function]\label{thm:soundLexBasic}
If there exists an input signal $\bu$ such that $\fitfal(\bu) = 0$, then the input constraints $\psi$ are satisfied and the specification $\varphi$ is falsified.
\end{theorem}

The proof directly comes from the definitions of $\fitfal$, $\trans_1$, and $\trans_2$, and the robustness definition of STL.

\subsection{Partially Simulation Free Lexicographic Method approach}\label{sec:lexMethodFalsVar}

In this section, we present a variation of the plain application of the lexicographic method presented in \S{}\ref{sec:lexMethodFals}. The current technique takes into account a particular feature of our problem: regarding the two objective functions in Eq.~\ref{eq:constraint} and Eq.~\ref{eq:robustness}, the computation of $\Robust{\bu}{\neg\psi}$ does not need system simulation, while computation of $\Robust{\M(\bu)}{\varphi}$ does. Since system simulation is the most time-consuming process (as we have already observed in~\S{}\ref{subsec:hillClimbingFalsification}), we adapt the $\fitfal$ function into a \emph{partially simulation free} version $\fitfalsf$ that avoids running simulations when $\psi$ is not satisfied, so saving time.

\begin{mydefinition}[Lexicographic fitness function $\fitfalsf$ for falsification]\label{def:lexicographicForFalsVar}
Let $f_1(\bu) \Defeq \Robust{\bu}{\neg\psi}$, and $f_2(\bu)\Defeq \Robust{\M(\bu)}{\varphi}$. The definition of the \emph{partially simulation free} global cost function for the constrained falsification problem is as follows:
\[\fitfalsf(\bu)=\left\{
\begin{array}{lll}
B\ceil{(B-1)\trans_1(f_1(\bu))}& & \quad\text{if } f_1(\bu) > 0\\% + (B-1)*1& & \quad\text{if } f_1(\bu)>0 \\
(B-1)\trans_2(f_2(\bu))& &\quad\text{otherwise}\\%f_1(\bu) \le 0 \\
\end{array}
\right.\]
\end{mydefinition}

Note that the only difference between Def.~\ref{def:lexicographicForFals} and Def.~\ref{def:lexicographicForFalsVar} is when the input constraints $\psi$ are not satisfied (first case): in this case, Def.~\ref{def:lexicographicForFalsVar} ignores the system specification in Eq.~\ref{eq:robustness} (so, no system simulation is performed), and thus $\fitfalsf(\bu)$ is only decided by the robustness of the input constraints, i.e., $\Robust{\bu}{\neg\psi}$; otherwise, it is the same as Def.~\ref{def:lexicographicForFals}. In the second case, we do not report the first operand of the sum that is 0 because the input constraints are satisfied.

%Note that the definition of $\fitfalsf$ still guarantees the order between inputs not satisfying the constraints and those satisfying them, i.e., the former still have higher fitness values than the latter.
Note that the definition of $\fitfalsf$ still guarantees the priorities between the two objective functions, i.e., inputs violating the input constraints still have higher fitness values than those satisfying them.

The soundness of the approach still holds, as stated in Thm.~\ref{thm:soundLexiMethVar}.

\begin{theorem}[Soundness of the $\fitfalsf$ fitness function]\label{thm:soundLexiMethVar}
If there exists an input signal $\bu$ such that $\fitfalsf(\bu) = 0$, then the input constraints $\psi$ are satisfied and the specification $\varphi$ is falsified.
\end{theorem}

The proof is similar to that of Thm.~\ref{thm:soundLexBasic}.

\section{Experimental Evaluation}\label{sec:experiments}

In order to evaluate the proposed techniques, we show their application to the benchmarks commonly used in the falsification community~\cite{ARCHCOMP19Falsification}. Specifically, we experimented them on 3 Simulink models, and 17 specifications to achieve comprehensive and reliable evaluation results. Note that the documents reporting the original Simulink models and temporal specifications do not provide any input constraints. Therefore, for each model, we identified some input constraints of different kinds, by using different logical and relational operators, and considering different input signals. The 3 Simulink models and their specifications are reported in Table~\ref{tab:spec}. The input constraints are reported in Table~\ref{tab:constraints}. 
\begin{table}[!tb]
\caption{Benchmarks of temporal specifications and input constraints in STL. Here, $\bw^t$ represents the \emph{t-shift} of $\bw$ (see Def.~\ref{def:robSemantics}) and  $\Delta_{t}(\bw)$ represents $\bw^{t}-\bw$}.
\begin{subtable}[t]{\textwidth}
\caption{Temporal specifications $\varphi$}
\label{tab:spec}
\centering
\resizebox{0.75\textwidth}{!}{%
\begin{tabular}{ccl}
\toprule
Model               & Spec. ID & Temporal specification in STL\\
\midrule
\multirow{13}{*}{AT} & AT1      &    $ \Box_{[0,30]}~(\speed < 120)$          \\ 
& AT2      &   $ \Box_{[0,30]}~(\gear = 3 \to \speed \ge 19)$          \\
& AT3      &     $ \Box_{[0,30]}~(\gear = 4 \to \speed \ge 35)$           \\
& AT4      &     $\neg (\Box_{[10,30]} ((50 < \speed) \land (\speed < 60))) $       \\
& AT5      &  $\neg (\Box_{[10,30]} ((53 < \speed) \land (\speed < 57))) $           \\
& AT6      &   $\Box_{[0,29]}(\speed<100) \lor \Box_{[29,30]}(\speed>75)$          \\
& AT7      &   $\Box_{[0,29]}(\speed<100) \lor \Box_{[29,30]}(\speed>70)$           \\
& AT8      &     $ \Box_{[0,30]}(\rpm < 4770 \lor \Box_{[0,1]}(\rpm > 1000))$         \\
& AT9      &    $ \Box_{[0,30]}(\rpm < 4770 \lor \Box_{[0,1]}(\rpm > 700))$         \\
& AT10     & $\Box_{[0, 30]} (\rpm < 3000) \to \Box_{[0, 20]} (\speed < 65)$            \\
& AT11     &     $\Box_{[0,10]}~(\speed<50) \lor \Diamond_{[0,30]}~(\rpm > 2520)$        \\
& AT12     &       $ \Box_{[0,26]}(\Delta_{4}(\speed) >40 \to \Delta_{4}(\gear) > 0)$       \\
& AT13     &       $ \Box_{[0,27]}(\Delta_{3}(\speed) >30 \to \Delta_{3}(\gear) > 0)$       \\\hline
\multirow{2}{*}{AFC} & AFC1     & $ \Box_{[11,50]} (\mu< 0.22) $        \\
& AFC2     &       $ \Box_{[11,50]} (\Diamond_{[0,10]} ( |\mu|<0.05 ))$      \\\hline
\multirow{5}{*}{NN} 	& &   $\nnreq\equiv\BoxOp{[0,16]}(\neg\closeref\to\reachref)$\\
&	& $\closeref \equiv |\pos-\reff|\le \alpha_1+\alpha_2\cdot|\reff| $\\
&	& $\reachref\equiv \DiaOp{[0,2]}(\BoxOp{[0,1]}(\closeref))$ \\
& NN1      & $\nnreq$ with $\alpha_1=0.003, \alpha_2 = 0.04$ \\
& NN2      &  $\nnreq$ with $\alpha_1=0.01, \alpha_2 = 0.03$          \\
\bottomrule
\end{tabular}
}
\end{subtable}
\begin{subtable}[t]{\textwidth}
\caption{Input Constraints $\psi$}
\label{tab:constraints}
\centering
\resizebox{0.75\textwidth}{!}{%
\begin{tabular}{ccl}
\toprule
Model        & Constr. ID & Constraint in STL \\
\midrule
\multirow{5}{*}{AT} & AT\_con1 & $\Box_{[0,30]}(\throttle = 0\lor \brake = 0)$\\
& AT\_con2 & $\Box_{[0,30]}(\throttle \le 20\lor \brake \le 50)$\\
& AT\_con3 & $\Box_{[0,30]}(\throttle >3\cdot \brake \lor \brake > 3\cdot\throttle)$\\
& AT\_con4 & $\Box_{[0, 24]}(\throttle > 70 \rightarrow \throttle^6 <10 )$\\
& AT\_con5 & $\Box_{[6,30]}(\throttle = 0\lor \brake = 0) \land \Box_{[0, 6]}(\brake = 0)$\\
\hline
\multirow{2}{*}{AFC} & AFC\_con1 & $\Box_{[0,50]}(\pedalangle \ge 50 \rightarrow \enginespeed>1000)$\\
& AFC\_con2 & $\Box_{[0, 20]}(\Delta_{10}(\pedalangle) \ge 0)$\\
\hline
\multirow{2}{*}{NN} & NN\_con1 & $\Box_{[0,12]}(\Delta_{6}(\reff) \ge 0)$\\
& NN\_con2 & $\Diamond_{[0,18]}(\reff > 2.5)$\\
\bottomrule
\end{tabular}
}
\end{subtable}
\end{table}

In the following, we provide a detailed description of the benchmarks.

The \emph{Automatic Transmission} (AT) model~\cite{HoxhaAF14} is a typical benchmark model in falsification. It has two input signals, $\throttle\in[0,100]$ and $\brake\in[0,325]$, and several output signals including $\speed$, $\rpm$, $\gear$, etc. Specifications AT1, \ldots, AT13 mainly concern safety of the system in different aspects. In the experiments, we consider 5 different input constraints, by considering both $\throttle$ and $\brake$, or only $\throttle$.

The \emph{Abstract Fuel Control} (AFC) model~\cite{JinDKUB14} takes two input signals, $\pedalangle$ and $\enginespeed$, and outputs the controller mode subject to $\pedalangle$, and a ratio $\mu$ reflecting the deviation of \emph{air-fuel-ratio} from its reference value. In our experiment, we set the range of $\pedalangle\in[8.8, 70]$ to keep the model in a \emph{normal} mode, and $\enginespeed\in[900, 1100]$ consistent with~\cite{JinDKUB14}. Specifications AFC1 and AFC2 reason about the related safety properties. We created two different input constraints, one constraining the value $\enginespeed$ w.r.t. the value of $\pedalangle$, and another one constraining the value of $\pedalangle$ over time.

The third benchmark model is based on MathWork's \emph{Neural Network controller} (NN) for a magnet system. Specifications NN1 and NN2 formalize the safety requirement about the position $\pos$ of the magnet w.r.t. its reference value $\reff$, which ranges over $[1,3]$. Since $\reff$ is the only input signal, we cannot reason about input constraints over different signals. Therefore, we just specified two input constraints over $\reff$: the first one requiring $\reff$ to be non-decreasing, and the second one requiring $\reff$ to be larger of 2.5 in at least one time point.

In the lexicographic method-based approaches proposed in~\S{}\ref{sec:lexMethodFals} and \S{}\ref{sec:lexMethodFalsVar}, we need to choose a proper base number $B$ and transformation functions $\trans_1$ and $\trans_2$ for the global cost function.
%In our experiment, we take the base number $B$ as a hyperparameter, so we try different selections of $B$ and analyze its influence in RQ3. In the experiments of Table~\ref{table:experimentalResults}, we select $10$ as the base number, the performance of which is beyond the average according to Table~\ref{table:baseNumber}.
Regarding $B$, we performed a preliminary experiment by comparing the performance of the approaches using different values of $B$: from the experiment described in RQ3, $B=10$ resulted to be one of two best settings (see Table~\ref{table:baseNumber}). Therefore, for the main experiments of the paper reported in Table~\ref{table:experimentalResults}, we used $10$ as base number $B$.
As for transformation functions $\trans_1$ and $\trans_2$, we need to determine \Rmax{\psi} and \Rmax{\varphi} in each case (see \S{}\ref{sec:lexMethodFals}).
%Deciding \Rmax{\psi} is not difficult: as the ranges of input signals are usually given, we can simply take the maximum value over the lengths of different input signals. Instead, deciding \Rmax{\varphi} needs some domain specific knowledge (although not too much) about the range of output signals. We handle this problem as follows: we take a small set of samplings and compute their robustness values; then, we enlarge the maximum value of the obtained robustness values by a reasonable factor, namely 1.5.
%Deciding \Rmax{\psi} is not difficult: as the ranges of input signals are usually given, we can simply take the maximum value over the lengths of different input signals. Instead, deciding \Rmax{\varphi} needs some domain specific knowledge (although not too much) about the range of output signals. 
We handle this problem as follows. We take a small set of samplings of the input space and compute their robustness values (both for the input constraint and the specification). Then, for the input constraints, we determine \Rmax{\psi} by multiplying the maximum value of the obtained robustness values by a reasonable factor, namely 1.5. For the specification, we determine \Rmax{\varphi} in a similar way.

In our experiments we use CMA-ES~\cite{AugerH05}, one of the state-of-the-art stochastic optimization algorithms for black box, as an implementation of hill-climbing optimization.

%\paragraph{Experiment platform}
The experiments use Breach version 1.2.13 on an Amazon EC2 c4.2xlarge instance (2.9 GHz Intel Xeon E5-2666 v3 Processor, 15~GB main memory).

\subsection{Evaluation}\label{sec:evaluation}

In order to evaluate our proposed approaches, we first check the performances of a state-of-the-art falsification tool (\breach) that does not consider input constraints during falsification; we name such unconstrained approach as \emph{Baseline Approach} (\BA). We run falsification using \BA over all the specifications reported in Table~\ref{tab:spec} with a timeout budget of 600 secs. In order to account for random variation of the approach, each experiment has been performed 30 times, by following guidelines of reporting results for randomized algorithms~\cite{Arcuri2011}. Table~\ref{table:resultsNoConstraints} reports the experimental results. 
\begin{table}[!tb]
\caption{Results of falsification without considering the input constraints  (FR: Falsification Rate (out of 30) -- CSR: Constraint Satisfaction Rate (out of falsifying inputs))}
\label{table:resultsNoConstraints}
\begin{subtable}[t]{\textwidth}
\caption{Automatic Transmission}
\centering
\resizebox{0.85\textwidth}{!}{%
\begin{tabular}{ccc|cc|cc|cc|cc|cc}
\toprule
&                         &                               & \multicolumn{10}{c}{CSR}\\
&                                &                               & \multicolumn{2}{c}{AT\_con1}   & \multicolumn{2}{c}{AT\_con2}    & \multicolumn{2}{c}{AT\_con3}    & \multicolumn{2}{c}{AT\_con4}    & \multicolumn{2}{c}{AT\_con5}    \\
\multicolumn{1}{l}{} & \multicolumn{1}{c}{FR (/30)} & \multicolumn{1}{c|}{time (s)} & \# & \multicolumn{1}{c|}{\%} & \# & \multicolumn{1}{c|}{\%} & \# & \multicolumn{1}{c|}{\%} & \# & \multicolumn{1}{c|}{\%} & \# & \multicolumn{1}{c}{\%} \\
\midrule
AT1                  & 30                             & 27.06                         & 1    & 3.3\%                    & 1    & 3.3\%                    & 1    & 3.3\%                    & 0    & 0                        & 0    & 0                        \\
AT2                  & 20                             & 29.3                          & 1    & 5\%                      & 7    & 35\%                     & 0    & 0                        & 20   & 100\%                    & 0    & 0                        \\
AT3                  & 12                             & 25.36                         & 0    & 0                        & 2    & 16.7\%                   & 1    & 8.3\%                    & 10   & 83.3\%                   & 0    & 0                        \\
AT4                  & 30                             & 41.06                         & 1    & 3.3\%                    & 3    & 10\%                     & 1    & 3.3\%                    & 26   & 86.7\%                   & 1    & 3.3\%                    \\
AT5                  & 28                             & 157.09                        & 0    & 0                        & 2    & 7.1\%                    & 2    & 7.1\%                    & 25   & 89.3\%                   & 0    & 0                        \\
AT6                  & 20                             & 96.3                          & 0    & 0                        & 0    & 0                        & 0    & 0                        & 0    & 0                        & 0    & 0                        \\
AT7                  & 18                             & 87.09                         & 0    & 0                        & 0    & 0                        & 0    & 0                        & 0    & 0                        & 0    & 0                        \\
AT8                  & 13                             & 58.88                         & 0    & 0                        & 1    & 7.6\%                    & 0    & 0                        & 0    & 0                        & 0    & 0                        \\
AT9                  & 13                             & 131.27                        & 0    & 0                        & 0    & 0                        & 0    & 0                        & 0    & 0                        & 0    & 0                        \\
AT10                 & 30                             & 46.04                         & 0    & 0                        & 3    & 10\%                     & 1    & 3.3\%                    & 30   & 100\%                    & 0    & 0                        \\
AT11                 & 23                             & 227.32                        & 0    & 0                        & 0    & 0                        & 0    & 0                        & 23   & 100\%                    & 0    & 0                        \\
AT12                 & 6                              & 50.6                          & 0    & 0                        & 1    & 16.7\%                   & 0    & 0                        & 0    & 0                        & 0    & 0                        \\
AT13                 & 21                             & 23.15                         & 0    & 0                        & 1    & 4.8\%                    & 0    & 0                        & 0    & 0                        & 0    & 0                        \\
\bottomrule 
\end{tabular}
}
\end{subtable}
\begin{subtable}[t]{0.48\textwidth}
\caption{Abstract Fuel Control}
\resizebox{\textwidth}{!}{%
\begin{tabular}{ccc|cc|cc}
\toprule
&            &          & \multicolumn{4}{c}{CSR}                      \\
&            &          & \multicolumn{2}{c}{AFC\_con1} & \multicolumn{2}{c}{AFC\_con2} \\
& FR (/30) & time (s) & \# & \multicolumn{1}{c|}{\%}            & \# & \%\\
\midrule
AFC1 & 30         & 44.79    & 8            & 26.7\%         & 1             & 3.3\%        \\
AFC2 & 6          & 211.82   & 0            & 0               & 0             & 0             \\
\bottomrule
\end{tabular}
}
\end{subtable}
\hspace{\fill}
\begin{subtable}[t]{0.45\textwidth}
\caption{Neural Network controller}
\resizebox{\textwidth}{!}{%
\begin{tabular}{ccc|cc|cc}
\toprule
&            &            & \multicolumn{4}{c}{CSR}                    \\
&            &            & \multicolumn{2}{c}{NN\_con1} & \multicolumn{2}{c}{NN\_con2} \\
& FR (/30) & time (s) & \# & \multicolumn{1}{c|}{\%}            & \# & \%\\
\midrule
NN1 & 20         & 163.57     &     0         & 0              &        8      & 40\%          \\
NN2 & 27         & 26.43      &     1         & 3.7\%         &        7      & 25.9\%       \\ \bottomrule
\end{tabular}
}
\end{subtable}
\end{table}
For each specification, it reports the \emph{falsification rate} (FR) as the number of experiments for which a falsifying input has been found, and the average execution \emph{time} over the successful executions. Moreover, for each input constraint $\psi$ reported in Table~\ref{tab:constraints}, we also check whether the found falsifying input satisfies (by chance) $\psi$: the Constraint Satisfaction Rate (CSR) reports the number and percentage of falsifying inputs that also satisfy the input constraints. FR informs us about the complexity of the falsification problem, and we will use it later in the experiments to see how handling the input constraints affects the falsification problem. Regarding CSR, we observe that, most of the times, the falsifying input violates the input constraint: in such a case, the \emph{falsifying area} of the input space is not strictly contained in the \emph{feasible area} satisfying the input constraints. In few cases, the input constraints are satisfied with a high percentage, meaning that there is a big overlap (if not proper inclusion in case of 100\%) between the falsifiable area and the feasible area.

Then, we run the three approaches proposed in the paper over all the benchmarks.\footnote{Technically, we modified the fitness evaluation of \breach to use the 3 new fitness functions.} We name as \constrEmb the Constraint Embedding approach presented in \S{}\ref{sec:stlBasedApproach}, as \lmBasic the approach based on Lexicographic Method presented in \S{}\ref{sec:lexMethodFals}, and as \lmAdv its modification presented in \S{}\ref{sec:lexMethodFalsVar}. Also in this case, all the experiments have been performed 30 times.

Table~\ref{table:experimentalResults} reports the experimental results.
\begin{table}[!tb]
\caption{Experimental results (FR: Falsification Rate)}
\label{table:experimentalResults}
\begin{subtable}[t]{\textwidth}
\caption{Automatic Transmission}
\centering
\resizebox{0.9\textwidth}{!}{%
\begin{tabular}{cc|cc|cc|cc|cc|cc}
\toprule
&            & \multicolumn{2}{c}{AT\_con1} & \multicolumn{2}{c}{AT\_con2} & \multicolumn{2}{c}{AT\_con3} & \multicolumn{2}{c}{AT\_con4} & \multicolumn{2}{c}{AT\_con5} \\
&            & FR (/30)          & time (s)         &FR (/30)          & time (s)         & FR (/30)          & time (s)           &FR (/30)          & time (s)         & FR (/30)          & time (s)          \\
\midrule
\multirow{3}{*}{AT1}  & \constrEmb & 18           & 78.62          & 26           & 64.05          & 14           & 88.43          & 13           & 367.26         & 15           & 114.72         \\
& \lmBasic     & 2            & 378.25         & 19           & 138.01         & 3            & 178.62         & 14           & 350.78         & 16           & 303.89         \\
& \lmAdv    & 0            & -             & 15           & 89.22          & 3            & 169.69         & 19           & 316.92         & 9            & 125.93         \\\hline
\multirow{3}{*}{AT2}  & \constrEmb & 5            & 85.19          & 18           & 44.91          & 23           & 62.21          & 22           & 24.7           & 10           & 59.57          \\
& \lmBasic     & 10           & 33.75          & 10           & 56.63          & 25           & 49.82          & 21           & 47.47          & 0            & -             \\
& \lmAdv    & 10           & 9.29           & 11           & 17.71          & 21           & 19.53          & 26           & 25.7           & 0            & -             \\\hline
\multirow{3}{*}{AT3}  & \constrEmb & 2            & 126.5          & 6            & 34.38          & 11           & 60.46          & 17           & 28.28          & 9            & 64.35          \\
& \lmBasic     & 6            & 38.05          & 5            & 49.92          & 11           & 83.93          & 16           & 15.81          & 0            & -             \\
& \lmAdv    & 6            & 26.49          & 7            & 24.72          & 14           & 29             & 17           & 27.24          & 0            & -             \\\hline
\multirow{3}{*}{AT4}  & \constrEmb & 23           & 136.14         & 30           & 73.7           & 9            & 80.81          & 30           & 35.37          & 23           & 143.71         \\
& \lmBasic     & 11           & 273.27         & 28           & 70.69          & 28           & 137.06         & 30           & 42.73          & 30           & 183.5          \\
& \lmAdv    & 12           & 132.63         & 28           & 175.28         & 26           & 86.96          & 30           & 42.98          & 23           & 74.72          \\\hline
\multirow{3}{*}{AT5} & \constrEmb & 21           & 260.97         & 28           & 195.83         & 8            & 278.95         & 30           & 156.36         & 13           & 259.86         \\
& \lmBasic     & 3            & 332.99         & 28           & 173.75         & 21           & 286.24         & 30           & 174.9          & 14           & 326.72         \\
& \lmAdv    & 5            & 239.26         & 28           & 175.28              & 25           & 180.69         & 30           & 134.08         & 17           & 243.24         \\\hline
\multirow{3}{*}{AT6}  & \constrEmb & 5            & 406.83         & 13           & 263.15         & 4            & 203.02         & 1            & 421.7          & 4            & 470.8          \\
& \lmBasic     & 1            & 594.79         & 5            & 405.46         & 5            & 317.91         & 1            & 395.75         & 0            & -             \\
& \lmAdv    & 0            & -             & 5            & 229.01         & 5            & 197.38         & 0            & -             & 0            & -             \\\hline
\multirow{3}{*}{AT7} & \constrEmb & 0            & -             &  0            & -              & 0            & -             & 4            & 465.65         & 0            & -             \\
& \lmBasic     & 0            & -             & 0            & -              & 5            & 351.57         & 2            & 528.73         & 0            & -             \\
& \lmAdv    & 0            & -             & 0            & -             & 2            & 203.09         & 2            & 395.26         & 0            & -             \\\hline
\multirow{3}{*}{AT8}  & \constrEmb & 7            & 362.45         & 8            & 241.13         & 1            & 450.03         & 0            & -             & 10           & 372.02         \\
& \lmBasic     & 7            & 184.5          & 6            & 86.59          & 1            & 176.33         & 0            & -             & 4            & 211.28         \\
& \lmAdv    & 5            & 99.62          & 9            & 72.49          & 1            & 26.84          & 0            & -             & 3            & 103.04         \\\hline
\multirow{3}{*}{AT9} & \constrEmb & 7            & 401.25         & 6            & 356.97         & 0            & -             & 0            & -             & 7            & 385.24         \\
& \lmBasic     & 10           & 182.46         & 9            & 70.64          & 1            & 105.46         & 0            & -             & 4            & 172.34         \\
& \lmAdv    & 3            & 75.76          & 12           & 72.27          & 0            & -             & 0            & -             & 5            & 108.18         \\\hline
\multirow{3}{*}{AT10}  & \constrEmb & 15           & 186.41         & 29           & 117.35         & 18           & 201.62         & 30           & 36.56          & 24           & 167.23         \\
& \lmBasic     & 7            & 133.63         & 25           & 149.34         & 25           & 182.6          & 30           & 28.28          & 17           & 81.18          \\
& \lmAdv    & 8            & 63.62          & 27           & 97.33          & 24           & 147.82         & 30           & 32.67          & 19           & 155.15         \\\hline
\multirow{3}{*}{AT11} & \constrEmb & 10           & 234.12         & 22           & 223.15         & 3            & 307.46         & 26           & 264.61         & 13           & 261.85         \\
& \lmBasic     & 2            & 184.39         & 22           & 220.04         & 1            & 554.55         & 21           & 260.33         & 1            & 51.71          \\
& \lmAdv    & 2            & 404.31         & 25           & 178.26         & 4            & 203.27         & 21           & 253.18         & 13           & 261.84         \\\hline
\multirow{3}{*}{AT12} & \constrEmb & 8            & 103.62         & 7            & 62.48          & 4            & 141.61         & 2            & 190.71         & 2            & 159.95         \\
& \lmBasic     & 9            & 147.38         & 15           & 89.01          & 11           & 118.55         & 1            & 166.02         & 8            & 149.39         \\
& \lmAdv    & 4            & 87.93          & 12           & 63.96          & 13           & 80.96          & 1            & 183.14         & 4            & 120.75         \\\hline
\multirow{3}{*}{AT13} & \constrEmb & 8            & 97.34          & 15           & 37.02          & 8            & 67.82          & 5            & 149.97         & 8            & 123.05         \\
& \lmBasic     & 16           & 147            & 15           & 61.32          & 15           & 116.86         & 10           & 74.4           & 7            & 108.82         \\
& \lmAdv    & 16           & 45.97          & 15           & 49.92          & 13           & 53.4           & 7            & 63.53          & 7            & 30.95            \\
\bottomrule
\end{tabular}
}
\end{subtable}
\begin{subtable}[t]{0.45\textwidth}
\centering
\caption{Abstract Fuel Control}
\resizebox{\textwidth}{!}{%
\begin{tabular}{cc|cc|cc}
\toprule
&            & \multicolumn{2}{c|}{AFC\_con1} & \multicolumn{2}{c}{AFC\_con2} \\
&            & FR (/30)          & time (s)         & FR (/30)          & time (s)         \\
\midrule
\multirow{3}{*}{AFC1} & \constrEmb &    25	   &  120.17         &       23       &     356.78          \\
& \lmBasic     &      29        &     56.32          &       29       &    	53.55           \\
& \lmAdv    &     29       &    49.03           &      29        &   46.89            \\\hline
\multirow{3}{*}{AFC2} & \constrEmb & 10           &  312.48             &    5          &   284.98            \\
& \lmBasic     &      11        &       350.47        &     10         &     139.01          \\
& \lmAdv   & 9           &    160.95           &      11        &    197.00          \\\bottomrule
\end{tabular}
}
\end{subtable}
\hspace{\fill}
\begin{subtable}[t]{0.45\textwidth}
\centering
\caption{Neural Network controller}
\resizebox{\textwidth}{!}{%
\begin{tabular}{cc|cc|cc}
\toprule
&            & \multicolumn{2}{c|}{NN\_con1} & \multicolumn{2}{c}{NN\_con2} \\
&            & FR (/30)          & time (s)         & FR (/30)          & time (s)         \\
\midrule
\multirow{3}{*}{NN1} & \constrEmb & 11           &   152.26            &      26        &     192.28          \\
& \lmBasic     &       16       &      181.65         &       24       &     139.79          \\
& \lmAdv    & 15           &       210.55        &  19            &      217.30         \\\hline
\multirow{3}{*}{NN2} & \constrEmb & 23          &       82.01        &      29        &  84.09             \\
& \lmBasic    &     19         &     66.45          &     30         &        67.99       \\
& \lmAdv    & 17           &    51.73           &      22        &   68.35       \\
\bottomrule    
\end{tabular}
}
\end{subtable}
\end{table}
Note that, by definition, all the approaches return falsifying inputs that respect the input constraints, i.e., CSR is always 100\% and so it is not reported. The table only reports FR and time.

We analyse the results using 3 research questions.

\researchquestion{Does constraint handling affect the falsifiability rate?}

First of all, we want to observe that, in most of the cases, FR of the three approaches is diminished w.r.t. that of \BA (i.e., \breach without constraint handling). This is expected, because almost all the falsifying inputs found by \BA do not satisfy the input constraints, and so our approaches correctly focus only on the feasible area. Note that, in the few cases in which also \BA had 100\% CSR (e.g., AT2 with input constraint AT\_con4), the falsification rate of the proposed approaches is the same as that of \BA, and sometimes even better. This holds also for cases in which CSR was high but not 100\% for \BA (e.g., AT3 with input constraint AT\_con4).

\researchquestion{How do the three proposed approaches perform?}

We are here interested in comparing the performance of the three proposed approaches. Regarding FR, in 11 out of 73 cases, the performances of the three approaches are the same. For the remaining 62 experiments, in 28 cases \constrEmb is strictly better or equal than the other two approaches. Although quite simple, \constrEmb can be effective in some cases. However, the lexicographic methods seem to be better on average.

Regarding \lmBasic and \lmAdv, in 28 cases they have the same FR, while in 24 cases \lmBasic is better than \lmAdv, and in 21 cases the other way round. This means that the optimization implemented by \lmAdv of not simulating the inputs that violate the input constraint, has a positive effect in some cases; however, when simulation is skipped, the objective function does not receive any contribution related to the robustness of the specification, and this may weaken the falsification ability of the approach.

Regarding the computation time when a falsifying input is found, \lmAdv is faster than \lmBasic in 47 cases out 61 (in which both approaches find a falsifying input). This confirms that \lmAdv does indeed speed up the process. However, there are some notable exceptions. For AT11 with input constraint AT\_con5, \lmAdv is much slower, but it has a much better falsification rate: this may be due to the fact that the time saved is used for exploring other inputs that turned out to be falsifiable and feasible (while \lmBasic, in 29/30 cases, timeouts without finding any falsifying input).

\researchquestion{Is there any influence in using different values for the base parameter in the lexicographic methods?}\label{rq:basenumber}

In \S{}\ref{sec:lexMethodFals}, we have described that the global cost function of a lexicographic method requires to define a base number $B$, that it is only required to be larger than 1. However, literature shows that different values of $B$ can affect the performance of the underlying optimization problems~\cite{Chang2015,pinchera2017lexicographic}. In this RQ, we investigate which is the effect of the choice of $B$ in our approaches. We selected 3 specifications of the AT benchmark (AT2, AT5, and AT13), and 2 input constraints (AT\_con2 and AT\_con3). For the six combinations, we run the two lexicographic methods \lmBasic and \lmAdv using 4 values for $B$, namely 5, 10, 100, and 1000. Results are reported in Table~\ref{table:baseNumber}.
\begin{table}[!tb]
\caption{Comparison of different values for base $B$ (FR: falsification rate)}
\label{table:baseNumber}
\setlength{\tabcolsep}{1pt}
\resizebox{\textwidth}{!}{%
\begin{tabular}{ccc|cccc|ccc|cccc|ccc|cccc}
\toprule
\multirow{2}{*}{}  & \multirow{2}{*}{}  & \multirow{2}{*}{base} & \multicolumn{2}{c}{AT\_con2} & \multicolumn{2}{c|}{AT\_con3} & \multirow{2}{*}{}   & \multirow{2}{*}{}   & \multirow{2}{*}{base} & \multicolumn{2}{c}{AT\_con2} & \multicolumn{2}{c|}{AT\_con3} & \multirow{2}{*}{}     & \multirow{2}{*}{}    & \multirow{2}{*}{base} & \multicolumn{2}{c}{AT\_con2} & \multicolumn{2}{c}{AT\_con3} \\
&                      &                       & FR (/30)         & time (s)         & FR (/30)        & time (s)      &                       &                      &                       & FR   (/30)      & time  (s)        & FR (/30)       & time (s)          &                       &                      &                       & FR (/30)        & time  (s)        & FR (/30)        & time (s)         \\
\midrule
\multirow{8}{*}{AT2} & \multirow{4}{*}{\lmBasic}  & 5                     & 9            & 45.44         & 26           & 41.04         & \multirow{8}{*}{AT5} & \multirow{4}{*}{\lmBasic}  & 5                     & 25           & 247.24        &      23        &     257.94          & \multirow{8}{*}{AT13} & \multirow{4}{*}{\lmBasic}  & 5                     & 13           & 63.51         & 14           & 68.09         \\
&                      & 10                    & 10           & 56.63         & 25           & 49.82         &                       &                      & 10                    & 28           & 173.75        & 21           & 286.24        &                       &                      & 10                    & 15           & 61.32         & 15           & 116.86        \\
&                      & 100                   & 15           & 34.78         & 22           & 43.22         &                       &                      & 100                   & 26           & 180.92        & 27           & 252.89        &                       &                      & 100                   & 16           & 53.73         & 10           & 133.94        \\
&                      & 1000                  & 13           & 33.33         & 20           & 46.38         &                       &                      & 1000                  & 25           & 261.10        & 25           & 267.52        &                       &                      & 1000                  & 11           & 90.65         & 6            & 182.43        \\\cline{2-7} \cline{9-14} \cline{16-21} 
& \multirow{4}{*}{\lmAdv} & 5                     & 9            & 16.16         & 24           & 13.49         &                       & \multirow{4}{*}{\lmAdv} & 5                     & 28           & 189.06        & 14           & 241.07        &                       & \multirow{4}{*}{\lmAdv} & 5                     & 11           & 34.06         & 13           & 51.13         \\
&                      & 10                    & 11           & 17.71         & 21           & 19.53         &                       &                      & 10                    & 28           & 175.28        & 25           & 180.69        &                       &                      & 10                    & 15           & 49.92         & 13           & 53.40         \\
&                      & 100                   & 16           & 26.07         & 25           & 20.80         &                       &                      & 100                   & 24           & 181.10        & 24           & 199.52        &                       &                      & 100                   & 14           & 46.60         & 12           & 84.43         \\
&                      & 1000                  & 13           & 30.07         & 24           & 26.53         &                       &                      & 1000                  & 26           & 174.37        & 28           & 191.11        &                       &                      & 1000                  & 10           & 72.59         & 10           & 117.91       \\
\bottomrule
\end{tabular}
}
\end{table}
We observe that there seems to be an effect on the falsification results. The two extreme cases of $B$ equal to 5 and to 1000 almost always produce the worst results, while the best results are distributed between the cases in which $B$ is 10 or 100. This is expected, as low values of $B$ produce more areas having flat robustness values (due to the combined use of the ceiling operator and $B$) for the input constraints and the specification: therefore, in this case, the search may not find the right direction. On the other hand, high values of $B$ generate a global cost function that prioritizes ``too much'' the first objective related to the input constraints, and a modification of the robustness of the specification has less effect on the global cost function (given a same value for the robustness of the input constraints).

\section{Related work}\label{sec:related}
Stochastic optimization-based falsification technique has drawn great many research attentions in recent years~\cite{FainekosP09,Annpureddy-et-al2011,AdimoolamDDKJ17,DeshmukhJKM15,KuratkoR14,Donze10,DonzeM10,DreossiDDKJD15,ZutshiDSK14,SilvettiPB17,falsificationTCAD2018,AkazakiLYDH18,ernstQEST2019,NghiemSFIGP10,Kato2019PRDC}, and becomes one of the most effective approaches to quality assurance of CPS products. Most of research efforts focus on developing or improving search techniques, and a lot of techniques were proposed to handle the ``exploration and exploitation'' trade-off, which is a core problem in search-based testing.  Notably some recent works~\cite{AkazakiLYDH18, falsificationTCAD2018, DeshmukhHJMP17} introduce advanced machine learning techniques into falsification, improving the effectiveness and efficiency substantially. A comparison of the state-of-the-art tools is given in~\cite{ARCHCOMP19Falsification}.

Our work bridges the gap between effectiveness and practicality of falsification, as few works consider the meaningfulness of  falsifying results. This problem was studied in~\cite{BarbotNFM2019}, where they use timed automata to formalize the input constraints and generate meaningful samplings. However, the proposed framework cannot be integrated into the state-of-the-art hill-climbing optimization-based falsification framework. Other examples include~\cite{JinDKUB14}, in which they mentioned an approach similar to our Constraint Embedding approach to handle an \emph{input profile}. Earlier works~\cite{NghiemSFIGP10} use sampling techniques so they can handle input constraints more complicated than bound constraints. 

The constrained optimization problem is one of the major research directions in the optimization community. However, a large amount of the research is based on white-box model.  Techniques on black-box models are more challenging as no derivative information is given. Genetic algorithm (GA) (or more generally, evolutionary algorithm (EA)) is a big branch of such techniques. A comprehensive list of literatures on handling constraints in GA is maintained~\cite{ConstraintsWeb}.
%One of the main used evolutionary algorithm for falsification is CMA-ES~\cite{AugerH05}, that it is also used by tool \breach and in our experiments. Methods handling constraints for CMA-ES exist~\cite{ArnoldH12,sakamoto2019adaptive,AtamnaAH16}, but they are not suitable for our needs because ...\todo{add some motivation}.

The constraint embedding approach builds  a specification  that predicates over both input and output signals.  The  approach in \cite{FerrereNDIK19} is tailored for handling safety properties having this combination of signals.  However, that approach is not  applicable to the constraint embedding approach which considers a different  class of properties.

\section{Conclusion and Future work}\label{sec:conclusion}

The paper presented three approaches for handling the input constraints in optimization-based falsification of hybrid systems. They implement, in different ways, a penalty method that adds a penalty factor to the fitness function that penalizes inputs that violate the input constraints. Experiments showed that each of the three approaches performs better in some cases. We believe that this depends on the relationship between the feasible area and the falsifying area of the input space. 
As future work, we plan to perform more detailed experiments in this direction to better characterize the strengths and weaknesses of the three approaches.
In particular, we  want to 
identify which constraints and/or specifications are better handled by a given method.

\bibliographystyle{abbrv}
\bibliography{biblio}

\end{document}